\documentclass[letterpaper,twocolumn,10pt]{article}
\usepackage{usenix2019_v3}

\usepackage{tikz}
\usepackage{amsmath}

\usepackage{color}
\usepackage{listings}
\usepackage{cite}  
\usepackage{authblk}
\usepackage{booktabs}
\usepackage{enumitem}
\usepackage{balance}
\usepackage{breakurl}

\usepackage[algoruled,linesnumbered]{algorithm2e}

\usepackage[firstpage]{draftwatermark}
\SetWatermarkText{\hspace*{6in}\raisebox{9.1in}{\includegraphics{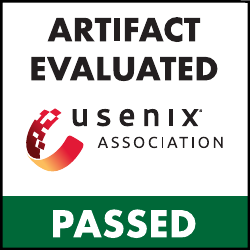}}}
\SetWatermarkAngle{0}

\usepackage{filecontents}

\def\Comment#1{\textbf{\textsl{\color{red}  $\langle\!\langle$#1$\rangle\!\rangle$}} }
\def\Comment#1{}

\def\System{Silhouette}
\def\SystemInvert{Silhouette-Invert}
\def\SSFI{SSFI}
\def\ISA{\mbox{ARMv7-M}}
\def\uXOM{{\it u}XOM}
\def\uRAI{$\mu$RAI}
\def\StoreTechnique{Store Hardening}
\def\storetechnique{store hardening}
\def\Storetechnique{Store hardening}
\def\StoreAbbreviation{SH}

\begin{document}

\date{}

\title{{\System}: Efficient Protected Shadow Stacks for Embedded Systems}


\author[1]{Jie Zhou}
\author[1]{Yufei Du}
\author[1]{Zhuojia Shen}
\author[1,2\thanks{Work done when the author was visiting
the University of Rochester.}]{Lele Ma}
\author[1]{John Criswell}
\author[3]{Robert J. Walls}
\affil[1]{University of Rochester}
\affil[2]{College of William \& Mary}
\affil[3]{Worcester Polytechnic Institute}

%
%

%
%

\maketitle

\begin{abstract}



Microcontroller-based embedded systems are increasingly used 
for applications that can have serious and immediate 
consequences if compromised---including 
automobile control systems, smart
locks, drones, and implantable medical devices.  Due to resource and
execution-time constraints, C is the primary language 
used for programming these devices.
Unfortunately, C is neither type-safe nor memory-safe, and
control-flow hijacking remains a prevalent threat.

This paper presents \emph{\System}: a compiler-based defense
that efficiently guarantees the integrity of return addresses, significantly 
reducing the attack surface for control-flow hijacking.
{\System} combines an incorruptible shadow stack for return addresses
with checks on forward control flow and memory protection to
ensure that all functions return to the correct dynamic caller.
To protect its shadow stack, {\System} uses
\emph{\storetechnique}, an efficient intra-address space isolation technique
targeting various ARM architectures that leverages
special store instructions found on ARM processors.

We implemented {\System} for the {\ISA} architecture, but our
techniques are applicable to other common embedded ARM architectures.
Our evaluation shows that {\System}
incurs a geometric mean of 1.3\% and 3.4\% performance overhead
on two benchmark suites.
Furthermore, we prototyped \emph{\SystemInvert},
an alternative implementation of {\System}, which incurs just 0.3\% and
1.9\% performance overhead, at the cost of a minor hardware change.
\end{abstract}


\section{Introduction}
\label{section:intro}

Microcontroller-based embedded systems are typically developed in C, meaning they suffer from the
same memory errors that have plagued general-purpose
systems~\cite{StackSmash:Phrack96,ROP:TISSEC12,EternalWar:Oakland13}.
Indeed, hundreds of vulnerabilities in embedded software have been
reported since 2017.\footnote{Examples include CVE-2017-8410, CVE-2017-8412, CVE-2018-3898,
CVE-2018-16525, CVE-2018-16526, and CVE-2018-19417.}
Exploitation of such systems can directly lead to physical
consequences in the real world. For example, the control system of a car is
crucial to passenger safety; the security of programs running
on a smart lock is essential to the safety of people's homes.  As these systems grow in
importance,\footnote{Both Amazon and
Microsoft have recently touted operating systems targeting
microcontroller-based embedded
devices~\cite{Microsoft:Sphere,FreeRTOS:Amazon}} their vulnerabilities become increasingly
dangerous~\cite{AutoAttack:BlackHat14,Thermostat:BlackHat14,IoT:DAC15}.


Past work on control-flow hijacking attacks highlights the need to
protect return addresses, even when the software employs other
techniques such as forward-edge control-flow integrity
(CFI)~\cite{OutOfControl:Oakland14,ROPDanger:USS14, StitchGadgets:USS14,
CFBending:USS15,LoseControl:CCS15}.
Saving return addresses on a separate shadow stack~\cite{SSSok:Oakland19}
is a promising approach, but shadow stacks themselves reside
in the same address space as the exploitable program and must be protected
from corruption~\cite{LoseControl:CCS15,SSSok:Oakland19}.
Traditional memory isolation that utilizes hardware privilege levels
can be adapted to protect the shadow stack~\cite{RECFISH:ECRTS19}, but
it incurs high overhead as there are frequent crossings between
protection domains (e.g., once for every function call).
Sometimes \emph{information hiding}
is used to approximate intra-address space isolation as it does
not require an expensive context switch.  In information hiding,
security-critical data structures are placed at a random location in memory
to make it difficult for adversaries to guess the
exact location~\cite{CPI:OSDI14}.
Unfortunately, information hiding is poorly suited to embedded systems
as most devices have a limited amount of memory that is directly
mapped into the address space---e.g., the board used in this work has just
384~KB of SRAM and 16~MB of SDRAM~\cite{STM32F469NI}.

This paper presents \emph{{\System}}: an efficient
write-protected shadow stack~\cite{SS:AsiaCCS15} system that guarantees that
a return instruction will always return to its dynamic legal destination.
To provide this guarantee, {\System} combines
a shadow stack, an efficient intra-address space isolation mechanism that
we call \emph{{\storetechnique}}, a Control-Flow Integrity~\cite{CFI:TISSEC09}
implementation to protect forward-edge control flow, and a corresponding
Memory Protection Unit (MPU) configuration to enforce memory access rules.
Utilizing the unprivileged store instructions on modern embedded ARM
architectures,
{\storetechnique}\footnote{{\uXOM}~\cite{uXOM:USS19} independently developed a
similar technique for implementing execute-only memory. We compare the
implementation differences between {\storetechnique} and that of {\uXOM} in
Section~\ref{section:impl:store}.}
creates a logical separation between the code and
memory used for the shadow stack and that used by application code.
Unlike hardware privilege levels, {\storetechnique}  does not require expensive
switches between protection domains. Also, unlike the probabilistic
protections of information hiding, protections based on {\storetechnique} are
hardware-enforced.
Further, the forward-edge control-flow protection prevents unexpected
instructions from being executed to corrupt the shadow stack or load
return addresses from illegal locations. Finally, the MPU configuration
enforces memory access rules required by {\System}.


We focus on the {\ISA} architecture~\cite{ARMv7M}
given the architecture's popularity and wide
deployment;
however, our techniques are also applicable to a wide range of
ARM architectures, including
\mbox{ARMv7-A}~\cite{ARMv7AR}
and the new \mbox{ARMv8-M Main Extension}~\cite{ARMv8M}.
We also explore an alternative, inverted
version of {\System} that promises significant performance improvements at
the cost of minor hardware changes; we call this version
\emph{{\SystemInvert}}. We summarize our contributions as follows:

\begin{itemize}
    \item
      We built a compiler and runtime system, {\System}, that leverages
      {\storetechnique} and coarse-grained CFI to provide embedded applications
      with efficient intra-address space isolation and a protected shadow
      stack.

    \item
    We have evaluated {\System}'s performance and code size overhead
    and found that {\System} incurs a geometric mean
    of 1.3\% and 3.4\% performance overhead, and a geometric mean of
    8.9\% and 2.3\% code size overhead on the CoreMark-Pro and the BEEBS
    benchmark suites, respectively. We also compare {\System} to two highly-related
    defenses: RECFISH~\cite{RECFISH:ECRTS19} and {\uRAI}~\cite{uRAI:NDSS20}.

  \item
    We prototyped and evaluated the {\SystemInvert} variant and
    saw additional improvements with an average performance overhead measured
    at 0.3\% and 1.9\% by geometric mean and code size overhead measured
    at 2.2\% and 0.5\%, again, on CoreMark-Pro and BEEBS.
\end{itemize}

In addition to the above contributions, we observe that
{\storetechnique} could be extended to protect other security-critical data,
making {\System} more flexible than other approaches.  For
example, {\System} could be extended to isolate the sensitive pointer
store for Code-Pointer Integrity (CPI)~\cite{CPI:OSDI14}.
Similarly, it could be used to protect
kernel data structures within an embedded operating system (OS) such as
Amazon FreeRTOS~\cite{FreeRTOS:Amazon}.


\section{{\ISA} Architecture}
\label{section:background}
%

Our work targets the {\ISA} architecture~\cite{ARMv7M}.
We briefly summarize the privilege and execution
modes, address space layout, and memory protection features of the {\ISA}.


\paragraph{Embedded Application Privilege Modes}
{\ISA} supports the execution of both \emph{privileged} and \emph{unprivileged}
code.  Traps, interrupts, and the execution of a
\emph{supervisor call} ({\tt SVC}) instruction switches the processor
from unprivileged mode to privileged mode.  Unlike server systems,
embedded applications often run in privileged mode.
Such applications also frequently use a \emph{Hardware Abstraction Layer} (HAL)
to provide a software interface to device-specific hardware.
HAL code is often generated by a manufacturer-provided tool
(e.g., HALCOGEN~\cite{TI:Halcogen}), is linked directly
into an application, and runs within its address space.

%


\paragraph{Address Space}
{\ISA} is a memory-mapped architecture, lacking support for virtual memory and
using a 32-bit address space.  While the exact layout varies between hardware, the
address space is generally divided into 8 sections.
The \texttt{Code} section holds code and
read-only data; it usually maps to an internal ROM or flash memory.
An \texttt{SRAM} section along with
two {\tt RAM} sections are used
to store runtime mutable data, e.g., the stack, heap, and globals.
The \texttt{Peripheral} and two \texttt{Device} regions map hardware timers
and I/O device registers.
The \texttt{System} region maps system control registers into the
processor's physical address space.

A security-critical subsection of \texttt{System} is the \emph{System Control Space},
which is used for important tasks such as system exception management. It
also contains the address space for the Memory Protection Unit (MPU)~\cite{ARMv7M}. Since
{\ISA} is a memory-mapped architecture, all of the security-critical registers,
such as MPU configuration registers, are also mapped to the \texttt{System}
region.

%
%
%

\paragraph{Memory Protection Unit}
An {\ISA}-based device can optionally have a Memory Protection Unit.
The MPU is a programmable memory protection component that enforces
memory access permissions~\cite{CortexM4,ARMv7M}.
The MPU allows privileged software to create a set of memory regions
which cover the physical address space; the permission bits on each
region dictate whether unprivileged and privileged memory accesses can
read or write the region.
The number of configurable MPU regions is implementation specific, e.g., the
target device in this paper supports $8$ regions\cite{STM32ProgManual}.  The memory regions
configured by the MPU do not need to exactly match the default memory regions
described in the {\bf Address Space} paragraph.
The size of each MPU-configured region varies from 32~bytes to 4~GB.


Currently, the MPU design makes several assumptions about how memory
access permissions are to be configured.  First, it assumes that
privileged software should have as many or more access rights to memory than
unprivileged code.  Consequently, the MPU cannot be configured to
give unprivileged code more access to a memory region than privileged
code.  Second, the MPU assumes that certain memory regions---e.g.,
the {\tt System} region---should not be executable, and it
prevents instruction fetches from these regions regardless of the MPU
configuration.  Third, the MPU design assumes that unprivileged code
should not be able  to reconfigure security-critical registers on the processor.
Therefore, the MPU will prevent unprivileged code from writing to memory
regions that include memory-mapped device registers, such as
those that configure the MPU.

\section{Threat Model and System Assumptions}
\label{section:threat}


While embedded code can be conceptually divided into application code,
libraries, kernel code, and the hardware abstraction layer,
there is often little distinction \emph{at runtime} between
these logical units.
Due to performance, complexity, and real-time considerations,
it is quite common for all
of this code to run in the same address space, without isolation,
and with the same privilege level~\cite{EPOXY:Oakland17,MINION:NDSS18,uXOM:USS19}.
For example, under
the default configuration of Amazon FreeRTOS (v1.4.7),
\emph{all} code runs as privileged in {\ISA}~\cite{FreeRTOS:Amazon}.
These embedded characteristics heavily inform our threat model and the design
decisions for {\System}.


Our threat model assumes a strong adversary that can
exploit a memory error in the \emph{application code} to create a
\emph{write-what-where} style of vulnerability. That is, the adversary can
attempt to write to any location in memory at any time.
The adversary's goal is to manipulate the control flow of a program by
exploiting the aforementioned memory error to overwrite  memory (e.g., a return
address).  Non-control data attacks~\cite{NonCtrlData:USS05,DOP:Oakland16}
are out of scope of this work.
Further, we assume the adversary has full knowledge of the memory
contents and layout; we do not rely on information hiding for protection.
Our threat model is consistent with past work on defenses against
control-flow hijacking.

We assume the target system runs a single bare-metal application statically
linked with all the library code and the hardware abstraction layer (HAL)---the
latter provides a device-specific interface to the hardware.
We assume the HAL is part of the Trusted Computing Base (TCB) and
is either compiled separately from the application code or annotated, allowing
{\System} to forgo transformations on the HAL that might preclude privileged
hardware operations.  Similarly, we assume that exception handlers are part of the TCB.
Further, we assume the whole binary runs in privileged mode for the reasons
mentioned previously.

Finally,  we assume the
target device includes a memory protection unit (or similar hardware mechanism)
for configuring coarse-grained memory permissions, i.e., {\System} is able
to configure read, write, and execute permissions for five regions
(summarized in Section~\ref{section:impl:mpu}) of the address space.

\section{Intra-Address Space Isolation}
\label{section:isolation}

\begin{figure*}[tb]
  \centering
  \includegraphics[scale=0.22]{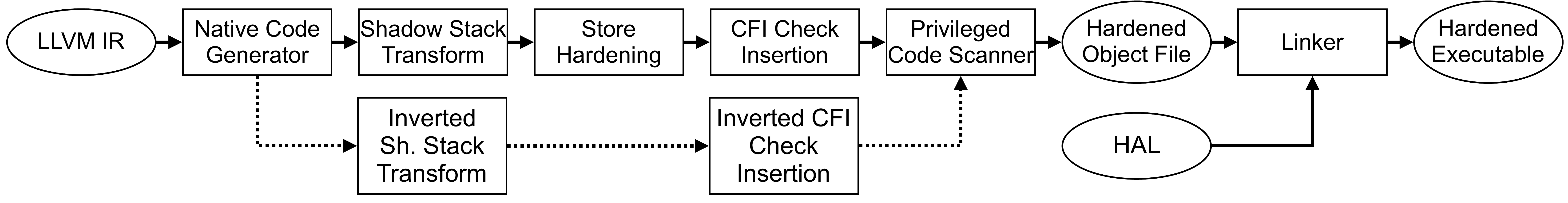}
  \caption{Architecture of {\System} and the {\SystemInvert} Variant}
  \label{fig:arch}
\end{figure*}

Many security enforcement mechanisms rely on intra-address space isolation to
protect security-critical data; in other words, the defenses are built on the
assumption that application code, under the influence of an attacker, cannot modify
security-critical regions of the address space.  For example, defenses with shadow
stacks~\cite{SSSok:Oakland19} need a safe region to store copies of return
addresses, and CPI~\cite{CPI:OSDI14} needs a
protected region of the address space to place its safe stack and
sensitive pointer store.
Complicating matters, defenses often intersperse accesses to the
protected region with regular application code; the former should be able to
access the protected region while the latter should not. Consequently,
existing mechanisms to switch between protection domains---e.g., system calls
between unprivileged and privileged mode---are often too inefficient for
implementing these security mechanisms for microcontroller-based embedded
systems.
Rather than incur the performance penalty of true memory isolation,
some defenses hide the security-critical
data structures at random locations in the address
space~\cite{EPOXY:Oakland17,CPI:OSDI14}.
Embedded systems have limited entropy sources for generating random numbers
and only kilobytes or megabytes of usable address space;
we do not believe hiding the shadow stack will be effective on such
systems.

We devise a protection method, \emph{{\storetechnique}},
for embedded ARM systems utilizing
unique features of a subset of ARM architectures~\cite{ARMv7M,ARMv7AR,ARMv8M},
including {\ISA}. These architectures provide
special unprivileged store instructions for storing 32-bit values ({\tt STRT}),
16-bit values ({\tt STRHT}), and 8-bit values ({\tt STRBT}).  When a program is
running in the processor's privileged mode, these store instructions are
treated as though they are executed in unprivileged mode, i.e., the processor
always checks the unprivileged-mode permission bits configured in the
MPU when executing an {\tt STRT}, {\tt STRHT}, or {\tt STRBT}
instruction regardless of whether the processor is executing in privileged or
unprivileged mode.  We leverage this feature to create two protection domains.
  One
\emph{unprivileged domain} contains regular
application code and only uses the unprivileged {\tt STRT}, {\tt STRHT}, and {\tt STRBT}
instructions for writing to memory. The second \emph{privileged domain} uses
regular (i.e., privileged) store instructions.
As code from both domains runs in the same, privileged, processor mode, this
method allows us to enforce memory isolation without costly context switching.

To completely isolate the data memory used by the unprivileged  and privileged
domains, two additional features are needed.  First, there needs to be a
mechanism to prevent unprivileged code from jumping into the middle of
privileged code; doing so could allow unprivileged code to execute a
privileged store instruction with arbitrary inputs.  We can use forward-edge CFI
checks to efficiently prevent such attacks.
Second, a trusted code scanner must ensure that the code contains no \emph{system
instructions} that could be used to modify important program state without the
use of a store instruction.  For example, an adversary could use the
{\tt MSR} instruction~\cite{ARMv7M} to change the value of the main or process stack
pointer registers ({\tt MSP} and {\tt PSP}, respectively), effectively
changing the location of the shadow stack and potentially moving it to an
unprotected memory region. We discuss a defense that leverages these techniques
in the next section.


\section{{\System} Design}
\label{section:design}

%
%


\emph{{\System}} is a compiler and run-time system that leverages our memory isolation scheme
to efficiently protect embedded
systems from control-flow hijacking attacks.  As Figure~\ref{fig:arch} shows,
{\System} transforms application code with four new compiler passes placed after
native code generation but before linking the hardened object code with the
hardware abstraction layer (HAL).
We also explore an alternative, inverted
version of these passes that promises significant performance improvements at
the cost of minor hardware changes; we call this version \emph{{\SystemInvert}}
(see Section~\ref{section:design:inversion}).
{\System}'s new compiler passes are as follows:

\begin{enumerate}
  \item{\textbf{Shadow Stack Transformation}:}
  The shadow stack transformation modifies the native code to save
  return values on a shadow stack and to use the return value stored in the
  shadow stack in return instructions.

  \item{\textbf{{\StoreTechnique}}:}
  The {\storetechnique} pass modifies all store instructions, except those used
  in the shadow stack instrumentation and
  Store-Exclusive instructions~\cite{ARMv7M}
  (see Section~\ref{section:design:store} for the reasons),
  to use variants that check the unprivileged-mode permission bits.

  \item{\textbf{CFI Transformation}:}
  The CFI transformation instruments indirect function calls
  and other computed branches (aside from returns) to ensure that
  program execution follows a pre-computed
  control-flow graph. Consequently, this instrumentation prevents the execution of
   gadgets that could, for example, be used to manipulate protected memory regions.

  \item{\textbf{Privileged Code Scanner}:}
  The privileged code scanner analyzes the native code prior to
  emitting the final executable to ensure that application code is free of
  privileged instructions that an adversary might seek to use to disable
  {\System}'s protections.

\end{enumerate}

In addition to the above transformations, {\System} employs mechanisms to
prevent memory safety errors from disabling the hardware features that
{\System} uses to provide its security guarantees. In the context of
{\ISA}, it means that the MPU cannot be reconfigured to allow unprivileged
accesses to restricted memory regions.
Also note that the HAL library is
not transformed with {\System} as it may contain I/O functions that need to
write to memory-mapped I/O registers that are only accessible to privileged
store instructions.  We also forbid inlining HAL functions into application
code.

Moreover, {\System} specially handles variable-length arrays on
the stack and {\tt alloca()} calls with argument values
that cannot be statically determined by the compiler. For these two types of
memory allocation, {\System} adopts the method from
SAFECode~\cite{SAFECode:PLDI06} and SVA~\cite{SVA:SOSP07} that promotes
the allocated data from stack to heap.
As Section~\ref{section:security:ss_integrity} explains,
such stack allocations (while rare in C code) can cause stack register spills,
endangering the integrity of the shadow stack.

\subsection{Shadow Stack}
\label{section:design:ss}

In unprotected embedded systems, programs store return addresses on the
stack, leaving return addresses open to corruption by an adversary.
To mitigate such attacks, some compilers transform code to use shadow
stacks.  A \emph{shadow stack}~\cite{SSSok:Oakland19} is a second
stack, stored in an isolated region of memory, on which a program
saves the return address.  Only the code that saves the return address
should be able to write to the shadow stack; it should be otherwise inaccessible
to other store instructions in the program.  If the shadow stack cannot
be corrupted by memory safety errors, then return addresses are not corrupted.
Furthermore, if the function epilogue uses the correct return address stored on the
shadow stack, then the function always returns to the correct dynamic
call site.

{\System}'s shadow stack transformation pass modifies each function's
prologue to save the return address on a shadow stack and each
function's epilogue to use the shadow stack return address on function return.
A special case to handle is {\tt setjmp}/{\tt longjmp}.
{\tt setjmp} saves the current execution context to a memory location
specified by its argument, and {\tt longjmp} recovers the saved context
from the specified memory location as if the execution was just returned
from a previous call to {\tt setjmp}.  Using {\tt setjmp}/{\tt longjmp},
a program is able to perform non-local indirect jumps that are
challenging to track by a shadow stack.  As few programs use
{\tt setjmp}/{\tt longjmp}, we refer
interested readers to Appendix~\ref{section:setjmp} which discusses how
{\System} supports these two functions.
Once the transformation is complete, the program uses
a shadow stack, but the shadow stack is not protected.  For that,
{\System} employs the {\storetechnique} pass and the CFI pass.

\subsection{Protection via {\StoreTechnique}}
\label{section:design:store}

{\System} leverages the MPU and the intra-address space isolation mechanism
described in Section~\ref{section:isolation} to efficiently protect the shadow
stack.  This protection is comprised of two parts.  First, during compilation,
{\System}'s \emph{{\storetechnique}} pass transforms all store instructions in
application code from privileged instructions to equivalent unprivileged store
instructions ({\tt STRT}, {\tt STRHT}, and {\tt STRBT}).  As discussed
previously, these unprivileged variants always check the MPU's
unprivileged-mode permission bits.  Second, when loading the program, {\System}
instrumentation configures the MPU so that the shadow stack is readable and
writeable in privileged mode but only readable in unprivileged mode.  This
ensures that store instructions executed in unprivileged mode and unprivileged
stores ({\tt STRT}, {\tt STRHT}, and {\tt STRBT}) executed in privileged mode
cannot modify values on the shadow stack. Together, these mechanisms ensure
shadow stack isolation, \emph{even if the entire program is executed in privileged mode}.

{\Storetechnique} transforms all stores within the application code
except for two cases. First, {\storetechnique} does not transform stores used
as part of {\System}'s shadow stack
instrumentation as they must execute as privileged instructions so that they can
write to the shadow stack.  The shadow stack pass marks all stores to the
shadow stack with a special flag, making them easily identifiable.
Second, {\storetechnique} cannot transform
atomic stores (Store-Exclusive~\cite{ARMv7M}) because
they do not have unprivileged counterparts. {\System} utilizes
Software Fault Isolation (SFI)~\cite{SFI:SOSP93} to prevent those stores
from writing to the shadow stack region.

As discussed in Section~\ref{section:threat}, {\System} does not
transform the HAL code; thus, the stores in the HAL code are
left unmodified. This is because the HAL contains hardware I/O and
configuration code that must be able to read and write the \texttt{System},
\texttt{Device}, and \texttt{Peripheral} memory regions.  To prevent
attackers from using privileged stores within the HAL code, {\System}
employs CFI as Section~\ref{section:design:cfi} explains.


\subsection{Forward Branch Control-Flow Integrity}
\label{section:design:cfi}

Shadow stacks protect the integrity of function
returns, but memory safety attacks can still corrupt
data used for forward-edge control flow branches, e.g., function pointers.  If
left unchecked, these manipulations would allow an attacker to redirect control
flow to anywhere in the program, making it trivial for the attacker to corrupt
the shadow stack with an arbitrary value or to load a return address from an
arbitrary location.  Consequently, {\System} must restrict the possible targets
of forward-edges to ensure return address integrity.

There are two types of forward branches: indirect function calls and
forward indirect jumps.  For the former, {\System} uses label-based
CFI checks~\cite{CFI:TISSEC09,CFISurvey:CSUR17} 
to restrict the set of branch targets and ensure that the remaining privileged
store instructions cannot be leveraged by an attacker to corrupt the shadow
stack.  {\System}-protected systems use privileged store instructions
only in the HAL library and in function prologues
to write the return address to the shadow stack.
The HAL library is compiled separately and
has no CFI labels in its code; even coarse-grained CFI ensures
that no store instructions within the HAL library can be exploited via
an indirect call (direct calls to HAL library functions are permitted
as they do not require CFI label checks).
For a function call, ARM processors automatically put the return address
in the {\tt lr} register. {\System}'s shadow stack transformation pass
modifies function prologues to
store {\tt lr} to the shadow stack. Label-based CFI guarantees an
indirect function call can only jump to the beginning of a function,
ensuring that attackers cannot use the function prologue to write
arbitrary values to the shadow stack.

There are three constructs in C that may cause a compiler to generate
forward indirect jumps: indirect tail function calls,
large {\tt switch} statements, and computed {\tt goto} statements
(``Label as Values'' in GNU's nomenclature~\cite{ComputedGoTo:GNU}).
{\System}'s CFI forces indirect tail function calls to jump to the beginning
of a function.
Restricting large {\tt switch} statements and computed {\tt goto} statements
is implementation-dependent. We explain how {\System} handles them
in Section~\ref{section:impl:cfi}.

\subsection{Privileged Code Scanner}
\label{section:design:scanner}

As {\System} executes all code within the processor's privileged mode,
{\System} uses a code scanner to  ensure the application code is free of
privileged instructions that could be used by an attacker to disable
{\System}'s protections. If the scanner detects such instructions, it presents
a message to the application developer warning that the security
guarantees of {\System} could
be violated by the use of such instructions. It is the application developer's
decision whether to accept the risk or modify the source code to avoid the use
of privileged instructions. 

On {\ISA}~\cite{ARMv7M}, there is only one privileged instruction that must be
removed: {\tt MSR} (Move to Special register from Register). One other, {\tt
CPS} (Change Processor State), must be rendered safe through hardware configuration.
Specifically, the {\tt MSR} instruction
can change special register values in ways that can subvert {\System}.
For example, MPU protections on the shadow stack could be bypassed by changing
the stack pointer registers ({\tt MSP} or {\tt PSP} on {\ISA})
to move the shadow stack to a memory region writeable by unprivileged code.
The {\tt CPS} instruction can change the execution priority,
and the MPU will elide protection checks if the current execution priority is
less than 0 and the {\tt HFNMIENA} bit in the MPU Control Register
({\tt MPU\_CTRL}) is
set to 0~\cite{ARMv7M}.  However, {\System} disables this feature by setting
the {\tt HFNMIENA} bit to 1, rendering the {\tt CPS} instruction
safe.  A third instruction, {\tt MRS} (Move to Register from Special
register), can read special registers~\cite{ARMv7M} but cannot be used to compromise the
integrity of {\System}.

Finally, as {\System} provides control-flow integrity,
an attacker cannot use misaligned instruction sequences to execute unintended
instructions~\cite{CFI:TISSEC09}.  Therefore,  a linear scan of the assembly
 is sufficient for ensuring
that the application code is free of dangerous  privileged instructions.

\subsection{Improvements with {\SystemInvert}}
\label{section:design:inversion}

Swapping a privileged store with a single equivalent unprivileged store
introduces no overhead. However, as
Section~\ref{section:impl:store} explains, {\System} must add additional
instructions when converting some privileged stores to unprivileged stores.
For example, transforming floating-point stores and stores with a large
offset operand adds time and space overhead.

However, we can minimize {\storetechnique}
overhead by \emph{inverting} the roles of hardware privilege modes.
Specifically, if
we can invert the permissions of the shadow stack region to disallow writes
from privileged stores but allow writes from unprivileged stores, then we can
leave the majority of store instructions unmodified. In other words, this
design would allow all stores (except shadow stack writes) to remain
unmodified, thereby incurring negligible space and time overhead for
most programs. We refer to this variant as \emph{{\SystemInvert}}.

\emph{{\SystemInvert}} is similar in design to ILDI~\cite{ILDI:DAC17}
which uses the Privileged Access Never (PAN) feature on
ARMv8-A~\cite{ARMv8A,ARMv8.1ANewFeature} to prevent privileged stores
from writing to user-space memory.  Unfortunately, the {\ISA} architecture
lacks PAN support and provides no way of configuring memory to be
writeable by unprivileged stores but inaccessible to privileged
stores~\cite{ARMv7M}.  We therefore reason about the potential performance
benefits using a prototype that mimics the overhead of a real
{\SystemInvert} implementation.  Section~\ref{section:impl:invert}
discusses two potential hardware extensions to {\ISA} to enable
development of {\SystemInvert}.

\subsection{Hardware Configuration Protection}
\label{section:design:protect-mpu-ivt}

As all code on our target system resides within a single address space and,
further, as {\System} executes application code in privileged mode to avoid
costly context switching, we must use both  the code transformations
described above and load-time hardware configurations to ensure that memory
safety errors cannot be used to reconfigure privileged hardware state. For
example, such state would include the interrupt vector table and
memory-mapped MPU configuration registers; on {\ISA}, most of this privileged hardware
state is mapped into the physical address space and can be modified
using store instructions~\cite{ARMv7M}.  If application code can write
to these physical memory locations, an adversary can reconfigure the MPU to make
the shadow stack writable or can violate CFI by changing the address of
an interrupt handler and then waiting for an interrupt to occur.
Therefore, {\System} makes sure that the MPU prevents these
memory-mapped registers from being writable by unprivileged store
instructions.  As Section~\ref{section:background} explains, the {\ISA}
MPU is automatically configured this way.

\section{Implementation}
\label{section:impl}

We implemented {\System} by adding three new
{\tt MachineFunction} passes to the
LLVM~9.0 compiler~\cite{LLVM:CGO04}: one that
transforms the prologue and epilogue code to use a shadow stack, one that
inserts CFI checks on all computed branches (except those used for returns),
and one that transforms stores into {\tt STRT}, {\tt STRHT}, or {\tt STRBT}
instruction sequences.
{\System} runs our new passes
after instruction selection and register allocation so that subsequent code
generator passes do not modify our instrumentation.
Finally, we implemented the privileged code scanner using a Bourne Shell
script which disassembles the final executable binary and searches for
privileged instructions.  Writing a Bourne shell script made
it easier to analyze code within inline assembly statements; such
statements are translated into strings within special instructions in the
LLVM code generator.
We measured the size of the
{\System} passes and code scanner using SLOCCount 2.26.
{\System} adds 2,416 source lines of C++ code to the code generator; the
code scanner is 95 source lines of Bourne shell code.

\subsection{Shadow Stack Transformation}
\label{section:impl:ss}

%
%
%


Our prototype implements a parallel shadow stack~\cite{SS:AsiaCCS15}
which mirrors the size and layout of the normal stack.
By using parallel shadow
stacks, the top of the shadow stack is always a constant offset from
the regular stack pointer.
Listing~\ref{fig:ss-prologue} shows the two instructions inserted
by {\System} in a function's prologue for our
STM32F469 Discovery board~\cite{STM32F469xx,STM32F469NI}.
The constant moved to the {\tt ip} register may vary across different devices
based on the available address space.
Note that the transformed prologue writes the return address into both
the regular stack and the shadow stack.


\begin{figure}[tb]
\begin{lstlisting}[caption={Instructions to Update the Shadow Stack},
    label={fig:ss-prologue}]
 mov.w ip, #0xe00000 // ip is the intra-procedure
                     // call scratch register
 str.w lr, [sp, ip]  // Save lr to mem[sp + ip]
\end{lstlisting}
\end{figure}

{\System} transforms the function epilogue to
load the saved return address to either {\tt pc} (program counter) or
{\tt lr}, depending on the instructions used in the original epilogue code.
The instructions added by the shadow stack transformation are marked
with a special flag so that a later pass (namely, the \emph{{\storetechnique}}
pass) knows that these instructions implement the shadow stack functionality.



{\System} also handles epilogue code within {\tt IT} blocks~\cite{ARMv7M}.
An {\tt IT} (short for If-Then)
instruction begins a block of up to 4 instructions
called an \emph{{\tt IT} block}.  An {\tt IT} block has a
condition code and a mask to control the conditional execution of the
instructions contained within the block. A compiler might generate an {\tt IT}
block for epilogue code if a function contains a
conditional branch and one of the branch targets contains
a {\tt return} statement.  For each such epilogue {\tt IT} block,
{\System} removes the {\tt IT} instruction, applies the epilogue
transformation, and inserts new {\tt IT} instruction(s) with the correct
condition code and mask to cover the new epilogue code.

\subsection{{\StoreTechnique}}
\label{section:impl:store}

{\System} transforms all possible variations of
regular stores to one of the three unprivileged store instructions:
{\tt STRT} (store word), {\tt STRHT} (store halfword), and
{\tt STRBT} (store byte)~\cite{ARMv7M}.
When possible, {\System} swaps the normal store with the
equivalent unprivileged store.  However, some
store instructions are not amenable to a
direct one-to-one translation.  For example, some store instructions use an
offset operand larger than the offset operand supported by the
unprivileged store instructions;  {\System} will insert additional
instructions to compute the target address in a register so that the
unprivileged store instructions can be used.  {\ISA} also supports instructions
that store multiple values to memory~\cite{ARMv7M}; {\System} converts
such instructions to multiple unprivileged store instructions.
For Store-Exclusive instructions~\cite{ARMv7M}, {\System} adds
two {\tt BIC} (bitmasking) instructions before the atomic store to
force the address operand to point into the
global, heap, or regular stack regions.

{\System} handles store instructions within {\tt IT}~\cite{ARMv7M}
blocks in a similar way to how it handles epilogue code within {\tt IT}
blocks.  If an {\tt IT} block has at least one store instruction,
{\System} removes the {\tt IT} instruction, applies {\storetechnique}
for each store instruction within the {\tt IT} block, and adds new
{\tt IT} instruction(s) to cover newly inserted instructions as well as
original non-store instructions within the old {\tt IT} block.
This guarantees {\storetechnique} generates semantically equivalent
instructions for every store in an {\tt IT} block.

{\System} sometimes adds code that must use a scratch register.
For example, when transforming floating-point store instructions,
{\System} must create code that moves the value from a floating-point
register to one or two integer registers because unprivileged store
instructions cannot access floating-point registers.
Our prototype uses
LLVM's {\tt LivePhysRegs} class~\cite{LivePhysRegs:LLVM} to find
free registers to avoid adding register spill code.
This optimization significantly reduces {\storetechnique}'s
performance overhead on certain programs; for example, we observed a reduction
from 39\% to 4.9\% for a loop benchmark.
Section~\ref{section:results:perf} presents detailed data of our experiments.

\paragraph{Comparison with {\uXOM}'s Store Transformation}
There are two major differences between {\System}'s implementation of
{\storetechnique} and the corresponding store transformation of
{\uXOM}~\cite{uXOM:USS19}.
First, {\System} performs
{\storetechnique} near the end of LLVM's backend pass pipeline
(after register allocation and right before the
{\tt ARMConstantIslandPass}~\cite{ARMConstantIslandPass:LLVM}).
We made this choice to avoid situations wherein later compiler passes
(potentially added by other developers) either
generate new privileged stores or transform instructions inserted by
{\System}'s shadow stack, {\storetechnique}, and CFI passes.
As mentioned above, {\System} avoids register spilling by
utilizing LLVM's {\tt LivePhysRegs} class to find free registers.
In contrast, {\uXOM} transforms store instructions prior to register
allocation to avoid searching for scratch registers.   As a consequence,
subsequent passes,
such as prologue/epilogue insertion or passes added by future developers,
must ensure that they do not add any new privileged store instructions.
Second, our {\storetechnique} pass transforms all privileged stores
(sans Store-Exclusives)
while {\uXOM} optimizes its transformation by
eliding transformation of certain stores
(such as those whose base register is {\tt sp}) when it is safe to do so.
The {\uXOM} optimization is safe when used with {\uXOM}'s security
policy but may not be safe if {\storetechnique} is used to enforce a
new security policy that does not protect the integrity of
the stack pointer register.
Implementing {\storetechnique} and optimization in a single pass
makes the compiler efficient.  However, by adhering to the
Separation of Concerns principle in
compiler implementation~\cite{PL.8:CC82}, our code is more easily reused:
to use {\storetechnique} for a new security policy, one simply changes
the compiler to run our {\storetechnique} pass and then implements any
optimization passes that are specific to that security policy.




\subsection{Forward Branch Control-Flow Integrity}
\label{section:impl:cfi}


%

\paragraph{Indirect Function Calls}
With link-time optimization enabled, {\System} inserts a CFI label at the
beginning of every address-taken function.  {\System} also inserts a 
check before each indirect call to ensure that the control flow transfers to a
target with a valid label.

Our prototype uses coarse-grained CFI checks, i.e., the prototype uses a single
label for all address-taken functions.  We picked {\tt 0x4600} for the CFI
label as it encodes the Thumb instruction {\tt mov r0, r0} and therefore
has no side effect when executed. With the addition of static call graph
analysis~\cite{DSA:PLDI07}, it is possible to extend the {\System} prototype to
use multiple labels with no increase in runtime overhead.


%
%
\begin{table}[ptb]
    \centering
    {\sffamily
    \footnotesize{
        \resizebox{\columnwidth}{!}{
        \begin{tabular}{@{}ll@{}}
            \toprule
            {\bf Code Pattern} & {\bf How {\System} Handles Them} \\
            \midrule
            Large {\tt switch} statement & Compiled to bounds-checked {\tt TBB} or {\tt TBH} \\
            Indirect tail function call & Restricted by CFI \\
            Computed {\tt goto} statement & Transformed to {\tt switch}
            statement \\
            \bottomrule
        \end{tabular}}
    }}
    \caption{C Code That May Be Compiled to Indirect Jumps}
    \label{table:indirect_jmp}
\end{table}

\paragraph{Forward Indirect Jumps}
Table~\ref{table:indirect_jmp} summarizes the three types of constructs
of C that may cause a compiler to generate a forward indirect jump and
how they are handled by {\System}.
The compiler may insert indirect jumps to implement large
{\tt switch} statements.  LLVM lowers large {\tt switch} statements into
PC-relative jump-table jumps using {\tt TBB} or {\tt TBH}
instructions~\cite{ARMv7M};
for each such instruction, LLVM places the jump table immediately after
the instruction and inserts a bounds check on the register holding the
jump-table index to ensure that it is within the bounds of the jump
table.  As jump-table entries are immutable and point to basic blocks
that are valid targets, such indirect jumps are safe.
Tail-call optimization transforms a function call
preceding a return into a jump to the
target function.  {\System}'s CFI checks ensure that tail-call
optimized indirect calls jump only to the beginning of a function.
The last construct that can generate indirect jumps is the computed
{\tt goto} statement.
Fortunately, LLVM compiles computed {\tt goto} statements into
{\tt indirectbr} IR instructions~\cite{LLVMLangRef}.  {\System} uses LLVM's
existing {\tt IndirectBrExpandPass}~\cite{IndirectBrExpand:LLVM}
to turn {\tt indirectbr} instructions into {\tt switch} instructions.
We can then rely upon LLVM's existing checks on {\tt switch} instructions,
described above, to ensure that indirect jumps generated from
{\tt switch} instructions are safe.
In summary, {\System} guarantees that no indirect jumps can jump to
the middle of another function.



\subsection{MPU Configuration}
\label{section:impl:mpu}

\begin{figure*}
  \centering
  \includegraphics[scale=0.50]{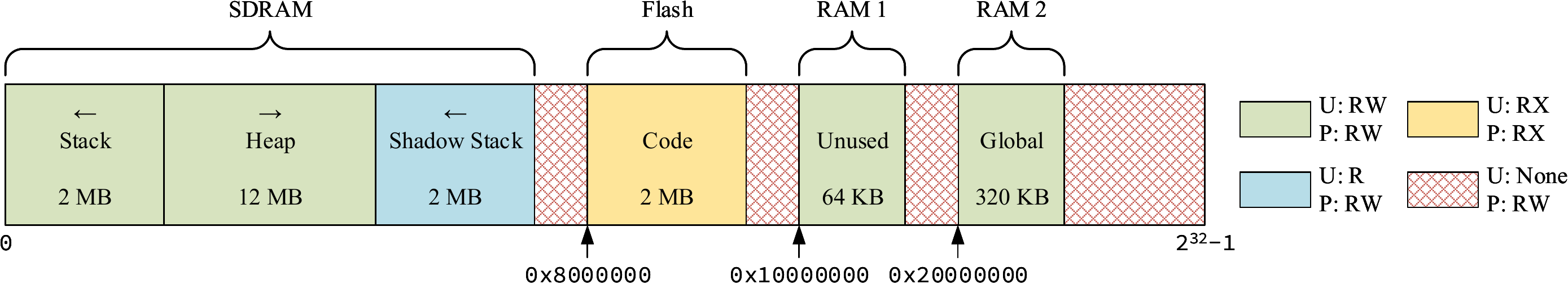}
  \caption{Address Space and MPU Configurations of {\System} on
  STM32F469 Discovery Board}
  \label{fig:addr_space_mpu}
\end{figure*}

Our prototype also includes code that configures the MPU
before an application starts.
Figure~\ref{fig:addr_space_mpu} shows the address space and the MPU
configuration for each memory region of a {\System}-protected system on
our STM32F469 Discovery board~\cite{STM32F469xx,STM32F469NI}.
{\System} uses five MPU regions to prevent unprivileged
stores from corrupting the shadow stack, program code, and hardware configuration.
First, {\System} sets the code region to be readable, executable, and
non-writable for both privileged and unprivileged accesses.  No other
regions are configured executable; this effectively enforces W$\oplus$X.
Second, {\System} configures the shadow stack region to be writable only
by privileged code.
All other regions of RAM are set to be readable and
writable by both privileged and unprivileged instructions.
Our prototype restricts the stack size to 2~MB; this should
suffice for programs on embedded
devices.\footnote{The default stack size of Android applications, including
both Java code and native code, is
only around 1~MB~\cite{AndroidStackSize}.}
Note that {\System} swaps the normal positions of the stack and the
heap to detect shadow stack overflow: a stack overflow will decrement
the stack pointer to point to the inaccessible region near the top of the
address space; a trap will occur when the prologue attempts to save
the return address there.
An alternative to preventing the overflow is to put an
inaccessible guard region between the stack and the heap;
however, it costs extra memory and an extra MPU configuration region.
Finally, {\System} enables the default background region which disallows
any unprivileged reads and writes to address ranges not covered by the above
MPU regions, preventing unprivileged stores from writing
the MPU configuration registers and the {\tt Peripheral}, {\tt Device}, and
{\tt System} regions.

\subsection{\SystemInvert}
\label{section:impl:invert}



Our {\SystemInvert} prototype assumes that the
hardware supports the hypothetical inverted-design described in
Section~\ref{section:design:inversion}, i.e., the MPU can be configured
so that the shadow stack is only writable in unprivileged mode.
We briefly propose two designs to change the hardware to support the memory
access permissions required by {\SystemInvert}.

One option is to use a reserved bit in the Application Program Status
Register ({\tt APSR})~\cite{ARMv7M} to support the PAN state mentioned
in Section~\ref{section:design:inversion}.
In ARMv8-A processors, PAN is controlled
by the {\tt PAN} bit in the Current Program Status Register
({\tt CPSR})~\cite{ARMv8A}. Currently, 24 bits of {\tt APSR} are
reserved~\cite{ARMv7M} and could be used for PAN on {\ISA}.

The second option is to add support to the MPU. In
{\ISA}, the permission configuration of each MPU region is defined
using three Access Permission (AP) bits in the
MPU Region Attribute and Size Register ({\tt MPU\_RASR})~\cite{ARMv7M}.
Currently, binary value {\tt 0b100} is reserved,
so one could map this reserved value to read
and write in unprivileged mode and no access in privileged mode, providing
support to the permissions required by {\SystemInvert} without changing the
size of AP or the structure of {\tt MPU\_RASR}.

In the {\SystemInvert} prototype, the function prologue writes the
return address to the shadow stack
using an unprivileged store instruction, and CFI uses regular store
instructions to save registers to the stack during label checks;
all other store instructions remain unchanged.
The MPU is also configured so that the shadow
stack memory region is writable in unprivileged mode, and other regions of RAM
are accessible only in privileged mode. As
configuring memory regions to be writable in unprivileged mode only would
require a hardware change, the {\SystemInvert} prototype instead configures
the shadow stack region to be writable
by both unprivileged and privileged stores.
We believe both of the potential hardware changes proposed above would
add negligible performance overhead.
Section~\ref{section:results} shows that {\SystemInvert}
reduces overhead considerably.

\subsection{Implementation Limitations}
\label{section:impl:limitation}

Our {\System} and {\SystemInvert} prototypes share a few limitations.
First, they currently do not transform inline assembly code.
The LLVM code generator represents inline assembly code within a C
source file as a special ``inline asm'' instruction with a
string containing the assembly code.  Consequently,
inline assembly code is fed directly into the
assembler without being transformed by {\tt MachineFunction} passes.
Fortunately, hand-written inline assembly code in applications is rare;
our benchmarks contain no inline assembly code.
Future implementations could implement {\storetechnique} within the assembler
which would harden stores in both compiler-generated and hand-written
assembly code.
Second, our current prototypes do not instrument the startup code
or the {\tt newlib} library~\cite{Newlib}.
These libraries are provided with our development board as pre-compiled
native code.  In principle, a developer can
recompile the startup files and {\tt newlib} from source code
to add {\System} and {\SystemInvert} protections.
Third, we have not implemented the ``stack-to-heap'' promotion (discussed
in Section~\ref{section:design}) for dynamically-sized stack data.
Only one of our benchmarks allocates a variable-length local array;
we manually rewrote the code to allocate the variable on the heap.
Lastly, we opted not to implement {\System}'s {\tt setjmp}/{\tt longjmp}
support, described in Appendix~\ref{section:setjmp}, as none of our benchmarks
use {\tt setjmp} and {\tt longjmp}.


\section{Security Analysis}
\label{section:security}

%

This section explains how {\System} hinders control-flow hijacking attacks.
We first discuss how {\System}'s protected shadow
stack, combined with the defenses on forward control-flow, ensure that
each return instruction transfers control back to its dynamic caller.
We then explain why these security mechanisms provide strong
protection against control-flow hijacking attacks.

\subsection{Integrity of Return Addresses}
\label{section:security:ss_integrity}

{\System} ensures that functions return control flow to their dynamic
callers when executing a return instruction by enforcing three
invariants at run-time:

\begin{itemize}[topsep=4pt,leftmargin=4pt]
\setlength\itemsep{0.25em}
  \item[] {\bf Invariant 1 (I1).} A function stores the caller's return
    address on the shadow stack, or never spills the return address in
    register {\tt lr} to memory.
\setlength\itemsep{0.25em}
    \item[] {\bf Invariant 2 (I2).}  Return addresses stored on the
      shadow stack cannot be corrupted.
    \item[] {\bf Invariant 3 (I3).} If a function stores the return
      address on the shadow stack, its epilogue will always retrieve
      the return address from the correct memory location in the shadow stack,
      i.e., the location into which its prologue stored
      the return address.
\end{itemize}


As the prologue and epilogue code use the stack pointer to compute the shadow
stack pointer, maintaining all the invariants requires maintaining the
integrity of the stack pointer.
Invariants {\bf I1} and {\bf I3} require the function prologue and
epilogue to keep the stack pointer within the stack region.
Additionally, for {\bf I3}, {\System} must ensure that the
stack pointer is restored to the \emph{correct} location on the stack
to ensure that the shadow stack pointer is pointing to the correct
return address. For {\bf I2}, besides being inside the stack region,
any function call's stack pointer must be guaranteed to stay lower
than its frame pointer; otherwise, the valid return addresses on the
shadow stack may be corrupted.

To maintain the invariants, {\System} prevents programs from
\emph{loading corrupted values into the stack pointer} by ensuring that
application code never spills and reloads the stack pointer to/from memory.
In particular, functions that have dynamically-sized stack allocations or that
allocate stack memory within a loop may trigger the code generator to
spill and reload the stack pointer.  As Section~\ref{section:design}
explains, {\System} promotes such problematic {\tt alloca}
instructions into heap allocations, ensuring that all
functions have constant-sized stack frames and therefore have no need
to spill the stack pointer.

The next issue is ensuring that the remaining fixed-size stack memory
allocations and deallocations cannot be used to violate the invariants.
To prevent \emph{stack overflow},
{\System} positions the regular stack at the bottom of the address space as
Figure~\ref{fig:addr_space_mpu} shows.
If a stack overflow occurs, the stack pointer will point to a location
near the top of the address space; if any function prologue
subsequently executes, it will attempt to write the return address
into an inaccessible location, causing a
trap that will allow the TCB to respond to the overflow.

To ensure that stack deallocation does not cause \emph{stack underflow},
{\System} ensures that deallocation frees the same amount of stack
memory that was allocated in the function prologue.  Several {\System}
features ensure this.  First, the checks on forward control flow
ensure that control is never transferred into the middle of a
function (as Section~\ref{section:impl:cfi} describes).
Second, if \textbf{I1}, \textbf{I2}, and \textbf{I3} hold prior to the
underflow,  then the shadow stack ensures that a function returns to the
correct caller, preventing mismatched  prologues and epilogues.
Finally, since the function prologue dominates all code in the function,
and since the function epilogue post-dominates all code in the function,
the epilogue will always deallocate the memory allocated in the prologue.

In summary, {\System} maintains {\bf I1} and {\bf I3} by ensuring that the
stack pointer stays within the stack region during the function prologue
and epilogue and that the epilogue will always deallocate stack memory
correctly.  {\System} also ensures that the stack pointer will always be
lower than the frame pointer, maintaining {\bf I2}.



\subsection{Reduced Attack Surface}
\label{section:security:attack_surface}

Recent work has shown the importance of protecting return addresses to
increase the precision, and thus strength, of CFI-based
defenses~\cite{OutOfControl:Oakland14,ROPDanger:USS14,
StitchGadgets:USS14,CFBending:USS15,LoseControl:CCS15}.
In particular, without a protected shadow stack or other mechanisms to ensure
the integrity of return addresses, CFI with static labels cannot ensure that a
function returns to the correct caller at runtime; instead, a function is
typically allowed to return to a \emph{set} of possible callers.  Attacks
against CFI exploit this imprecision.


Most attacks against CFI target programs running on general-purpose
systems. Some attacks exploit features specific to certain platforms,
and it is not clear if they can be ported to attack embedded devices.
For example, Conti et al.~\cite{LoseControl:CCS15} showed how to
corrupt return addresses saved by unprotected context switches on
Windows on 32-bit x86 processors.
However, many attacks involve generic code patterns that can likely be
adapted to attack CFI-protected programs on embedded systems.
We now discuss generic control-flow hijacking code patterns
discovered by recent
work~\cite{OutOfControl:Oakland14,ROPDanger:USS14,StitchGadgets:USS14,CFBending:USS15}.
As we discuss below, {\System} is robust against these attacks.

G\"{o}ktas et al.~\cite{OutOfControl:Oakland14} evaluated
the effectiveness of coarse-grained CFI that allows two types of
gadgets: {\it Call-site} (CS) gadgets that start after a function call
and end with a return, and {\it Entry-point} (EP) gadgets that start
at the beginning of a function and end with any indirect control transfer.
CS gadgets are a result of corrupted return addresses, and EP gadgets
stem from corrupted function pointers or indirect jumps if
the CFI policy does not distinguish indirect calls and jumps.
The authors proposed four methods of chaining the gadgets: CS to CS
(i.e., return-oriented programming),
EP to EP (call-oriented programming), EP to CS, and CS to EP.
Three of these four methods require a corrupted return address.
Their proof-of-concept exploit uses both types of the gadgets.
Similarly, Carlini et al.~\cite{ROPDanger:USS14} and
Davi et al.~\cite{StitchGadgets:USS14} showed how to chain
{\it call-preceded} gadgets (instruction sequences starting
right after a call instruction) to launch code-reuse attacks against CFI.
As {\System} prevents return address corruption, only attacks that chain
EP gadgets are possible.

Carlini et al.~\cite{CFBending:USS15} also demonstrated the
weaknesses of CFI and emphasized the importance of a shadow stack.
They proposed a {\it Basic Exploitation Test} (BET)---i.e., a minimal program
for demonstrating vulnerabilities---to quickly test the
effectiveness of a CFI policy.   Their
work identifies five dangerous gadgets that allow arbitrary reads, writes,
and function calls in the BET under a coarse-grained CFI policy.
However, all of these are call-preceded gadgets, and
{\System}'s protected shadow stack stymies call-preceded gadgets.

Additionally, Carlini et al.~\cite{CFBending:USS15} demonstrated a
fundamental limitation  of CFI
defenses when used \emph{without} another mechanism to provide return address
integrity.  Specifically, they showed that even fully-precise static CFI cannot
completely prevent control-flow hijacking attacks, concluding that, regardless
of the precision of the computed call graph, protection for return addresses is
needed.

In summary, with the protection of {\System}, control-flow hijacking
attacks are restricted to only call-oriented programming. Although
there are still potential dangers~\cite{ControlJujutsu:CCS15},
{\System} significantly reduces the control-flow hijacking attack surface
for embedded programs.

\section{Experimental Results}
\label{section:results}

Below, we evaluate the performance and code size overhead of our {\System} and
{\SystemInvert} prototypes. We also compare {\System} to an
orthogonal approach, \emph{SSFI}, which uses Software Fault Isolation (SFI),
instead of {\storetechnique}, to isolate the shadow
stack from application code. In summary, we find that {\System} and
{\SystemInvert} incur low runtime overhead (1.3\% and 0.3\% on average for
CoreMark-Pro, respectively) and small increases in code size (8.9\% and
2.2\%, respectively).
In addition, we compare {\System} with the two most closely related
defenses, RECFISH~\cite{RECFISH:ECRTS19} and {\uRAI}~\cite{uRAI:NDSS20};
they both protect return addresses of programs running on
microcontroller-based embedded devices but leverage different mechanisms than
{\System}.


\subsection{Methodology}

We evaluated {\System} on an STM32F469 Discovery
board~\cite{STM32F469xx,STM32F469NI}
that can run at speeds up to 180~MHz. The board encapsulates
an ARM Cortex-M4
processor~\cite{CortexM4} and has
384~KB of SRAM (a 320~KB main SRAM region and a 64~KB CCM RAM region),
16~MB of SDRAM,
and 2~MB of flash memory.
As some of our benchmarks allocate megabytes of memory,
we use the SDRAM as the main memory for all programs;
global data remains in the main SRAM region.

We used unmodified Clang 9.0 to compile all benchmark
programs as the baseline, and we compare this baseline with
programs compiled by {\System}, {\SystemInvert}, and SSFI for
performance and code size overhead.
We also measured the overhead incurred for each benchmark
program when transformed with
only the shadow stack (SS) pass,
only the {\storetechnique} ({\StoreAbbreviation}) pass,
and  only the CFI pass.
For all experiments, we used the standard {\tt -O3} optimizations,
and we used LLVM's {\tt lld} linker with the {\tt -flto} option
to do link-time optimization.

As {\SystemInvert} requires a hardware enhancement for a fully-functional
implementation, the numbers we present here are an estimate of
{\SystemInvert}'s performance. However, as
Sections~\ref{section:design:inversion} and~\ref{section:impl:invert}
discuss, the
hardware changes needed by {\SystemInvert} should have minor impact on
execution time and no impact on code size. Therefore,
our evaluation of the {\SystemInvert} prototype should provide an
accurate estimate of its performance and memory overhead.

We discuss
the implementation of SSFI and compare it with {\System} and {\SystemInvert} in
Section~\ref{section:results:sfi}.


%
%
\begin{table}[ptb]
\centering
{\sffamily
\footnotesize{
\resizebox{\columnwidth}{!}{%
\begin{tabular}{@{}lrrrrrrr@{}}
\toprule
  & {\bf Baseline} & {\bf SS} & {\bf {\StoreAbbreviation}} & {\bf CFI} & {\bf Silhou-} & {\bf Invert} & {\bf {\SSFI}} \\
  & {\bf (ms)} & {\bf ($\times$)} & {\bf ($\times$)} & {\bf ($\times$)} & {\bf ette ($\times$)} & {\bf ($\times$)} & {\bf ($\times$)}\\
\midrule
  cjpeg-rose7-... & 12,765 & 1.002 & 1.004 & 1.001 & 1.006 & 1.003 & 1.041 \\
  core & 137,385 & 1.013 & 1.002 & 1.000 & 1.017 & 1.015 & 1.024 \\
  linear\_alg-... & 18,278 & 1.000 & 1.010 & 1.000 & 1.010 & 1.000 & 1.015 \\
  loops-all-... & 35,241 & 1.000 & 1.049 & 1.000 & 1.049 & 1.000 & 1.016 \\
  nnet\_test & 222,461 & 1.000 & 1.013 & 1.000 & 1.013 & 1.000 & 1.023 \\
  parser-125k & 9,985 & 1.004 & 1.001 & 1.001 & 1.005 & 1.004 & 1.009 \\
  radix2-big-64k & 17,270 & 1.000 & 1.007 & 1.000 & 1.007 & 1.000 & 1.019 \\
  sha-test & 40,725 & 1.002 & 1.005 & 0.999 & 1.007 & 1.005 & 1.046 \\
  zip-test & 19,955 & 1.000 & 1.000 & 1.000 & 1.001 & 1.000 & 1.006 \\
\midrule
  {\bf Min} & 9,985 & 1.000 & 1.000 & 0.999 & 1.001 & 1.000 & 1.006 \\
  {\bf Max} & 222,461 & 1.013 & 1.049 & 1.001 & 1.049 & 1.015 & 1.046 \\
  {\bf Geomean} & --- & 1.002 & 1.010 & 1.000 & 1.013 & 1.003 & 1.022 \\
\bottomrule
\end{tabular}
}}}
\caption{Performance Overhead on CoreMark-Pro}
\label{table:coremarkpro_perf}
\end{table}

%
%
\begin{table}[ptb]
\centering
{\sffamily
\footnotesize{
\resizebox{\columnwidth}{!}{%
\begin{tabular}{@{}lrrrrrrr@{}}
\toprule
  & {\bf Baseline} & {\bf SS} & {\bf {\StoreAbbreviation}} & {\bf CFI} & {\bf Silhou-} & {\bf Invert} & {\bf {\SSFI}} \\
  & {\bf (ms)} & {\bf ($\times$)} & {\bf ($\times$)} & {\bf ($\times$)} & {\bf ette ($\times$)} & {\bf ($\times$)} & {\bf ($\times$)}\\
\midrule
  bubblesort & 2,755 & 1.001 & 1.247 & 1.000 & 1.248 & 1.000 & 1.510 \\
  ctl-string & 1,393 & 1.015 & 1.011 & 0.999 & 1.027 & 1.016 & 1.035 \\
  cubic & 28,657 & 1.002 & 1.002 & 1.000 & 1.002 & 1.001 & 1.005 \\
  dijkstra & 40,580 & 1.002 & 1.001 & 1.000 & 1.003 & 1.002 & 1.117 \\
  edn & 2,677 & 1.000 & 1.004 & 1.000 & 1.004 & 1.000 & 1.058 \\
  fasta & 16,274 & 1.000 & 1.000 & 1.000 & 1.000 & 1.000 & 1.001 \\
  fir & 16,418 & 1.000 & 1.000 & 1.000 & 1.000 & 1.000 & 1.021 \\
  frac & 8,846 & 1.000 & 1.003 & 1.000 & 1.000 & 1.000 & 1.009 \\
  huffbench & 46,129 & 1.000 & 1.005 & 1.000 & 1.005 & 1.000 & 1.017 \\
  levenshtein & 7,835 & 1.005 & 1.019 & 1.000 & 1.207 & 1.186 & 1.248 \\
  matmult-int & 5,901 & 1.000 & 1.011 & 1.000 & 1.012 & 1.000 & 1.048 \\
  nbody & 124,578 & 1.000 & 0.997 & 1.000 & 0.997 & 1.000 & 1.003 \\
  ndes & 1,938 & 1.010 & 1.008 & 1.000 & 1.016 & 1.011 & 1.039 \\
  nettle-aes & 7,030 & 1.000 & 1.003 & 1.000 & 1.003 & 1.000 & 1.111 \\
  picojpeg & 43,010 & 1.037 & 1.057 & 0.997 & 1.098 & 1.037 & 1.380 \\
  qrduino & 43,564 & 1.000 & 1.036 & 1.000 & 1.036 & 1.000 & 1.072 \\
  rijndael & 78,849 & 1.001 & 1.008 & 1.000 & 1.008 & 1.005 & 1.146 \\
  sglib-dllist & 1,327 & 1.001 & 1.006 & 1.000 & 1.007 & 1.001 & 1.268 \\
  sglib-listins... & 1,359 & 1.001 & 1.000 & 1.000 & 1.001 & 1.001 & 1.054 \\
  sglib-listsort & 1,058 & 1.001 & 0.999 & 1.000 & 1.000 & 1.001 & 1.233 \\
  sglib-queue & 2,135 & 1.000 & 1.029 & 1.000 & 1.030 & 1.000 & 1.122 \\
  sglib-rbtree & 7,802 & 1.092 & 1.017 & 1.000 & 1.110 & 1.093 & 1.157 \\
  slre & 4,163 & 1.031 & 1.013 & 1.000 & 1.045 & 1.035 & 1.112 \\
  sqrt & 55,894 & 1.000 & 1.002 & 1.000 & 1.006 & 1.002 & 1.002 \\
  st & 20,036 & 1.002 & 1.002 & 1.002 & 1.002 & 1.002 & 1.008 \\
  stb\_perlin & 3,168 & 1.073 & 1.052 & 1.000 & 1.049 & 1.073 & 1.045 \\
  trio-sscanf & 1,335 & 1.037 & 1.006 & 1.022 & 1.073 & 1.063 & 1.115 \\
  whetstone & 97,960 & 1.000 & 1.001 & 1.000 & 1.001 & 1.000 & 1.002 \\
  wikisort & 160,307 & 1.011 & 1.013 & 1.016 & 1.039 & 1.029 & 1.180 \\
\midrule
  {\bf Min} & 1,058 & 1.000 & 0.997 & 0.997 & 0.997 & 1.000 & 1.001 \\
  {\bf Max} & 160,307 & 1.092 & 1.247 & 1.022 & 1.248 & 1.186 & 1.510 \\
  {\bf Geomean} & --- & 1.011 & 1.018 & 1.001 & 1.034 & 1.019 & 1.102 \\
\bottomrule
\end{tabular}
}}}
\caption{Performance Overhead on BEEBS}
\label{table:beebs_perf}
\end{table}

\subsection{Benchmarks}
\label{section:results:becnhmark}

We chose two benchmark suites for our evaluation:
CoreMark-Pro~\cite{CoreMarkPro:EEMBC} and
BEEBS~\cite{BEEBS:arXiv}.  The
former is the de facto industry standard benchmark for embedded processors;
the latter has been used in the evaluation of other embedded
defenses~\cite{RECFISH:ECRTS19,EPOXY:Oakland17,uXOM:USS19}.


\paragraph{CoreMark-Pro}
The CoreMark-Pro~\cite{CoreMarkPro:EEMBC} benchmark suite
is designed for both low-end microcontrollers and high-end multicore
processors. It includes five integer workloads (including
JPEG compression and SHA-256) and four floating-point
workloads such as fast Fourier transform (FFT) and a neural network benchmark.
One of the workloads is
a more memory-intense version of the original CoreMark
benchmark~\cite{CoreMark:EEMBC}; note, ARM recommends the use of the original
CoreMark benchmark to test Cortex-M processors~\cite{CoreMark:ARM}.
We used commit {\tt d15927b} of the CoreMark-Pro
repository on GitHub.

The execution time of CoreMark-Pro is reported by benchmarks themselves,
which is by calling {\tt HAL\_GetTick()}~\cite{HALLib} to mark the start
and the end of benchmark workload execution and printing out the time
difference in milliseconds.  We added code before the main function
starts to initialize the HAL, set up the clock speed, configure the MPU,
and establish a serial output.  We run each CoreMark-Pro benchmark in
different number of iterations so that the baseline execution time is
between 5~to 500~seconds.

\paragraph{BEEBS}
The BEEBS benchmark suite~\cite{BEEBS:arXiv} is designed for measuring
the energy consumption of
embedded devices.  However, it is also useful for evaluating
performance and code size overhead because it includes a wide range of
programs, including a benchmark based on the Advanced Encryption Standard (AES),
integer and floating-point matrix multiplications, and an advanced
sorting algorithm.

The major drawback of BEEBS is that many of its programs either are too small
or process too small inputs, resulting in insufficient execution time.  For
example, {\tt fibcall} is intended to compute the 30th Fibonacci number, but
Clang computes the result during compilation and returns a constant
directly.
To account for this issue, we exclude programs with a baseline execution time
of less than one second with 10,240 iterations.
We also exclude {\tt mergesort} because it failed the
{\tt verify\_benchmark()} check when compiled with unmodified Clang.
For all the other programs, all of our transformed versions passed
this function, if it was implemented.
We used commit {\tt 049ded9} of the BEEBS repository on GitHub.

To record the execution time of an individual BEEBS benchmark,
we wrapped 10,240 iterations of benchmark workload execution with calls
to {\tt HAL\_GetTick()}~\cite{HALLib} and added code to print out the
time difference in milliseconds.  We also did the same initialization
sequence for each BEEBS benchmark as we did for CoreMark-Pro.



\subsection{Runtime Overhead}
\label{section:results:perf}

Tables~\ref{table:coremarkpro_perf} and~\ref{table:beebs_perf}
show the performance overhead that {\System} and {\SystemInvert}
induce on CoreMark-Pro and BEEBS, respectively; overhead is
expressed as execution time normalized to the baseline.  The \textbf{SS}
column shows the overhead of just the shadow stack transformation,
\textbf{{\StoreAbbreviation}} shows the overhead induced when only {\storetechnique} is
performed, and \textbf{CFI} shows the overhead of the CFI checks on
forward branches.  The \textbf{\System} and \textbf{Invert} columns show
the overhead of the complete {\System} and {\SystemInvert} prototypes,
respectively.  The \textbf{SSFI} column denotes overhead incurred by a
version of {\System} that uses Software Fault Isolation (SFI) in place
of {\storetechnique}; Section~\ref{section:results:sfi} describes that
experiment in more detail.

\paragraph{{\System} Performance}
As Tables~\ref{table:coremarkpro_perf} and~\ref{table:beebs_perf} show,
{\System} incurs a geometric mean overhead of only
\emph{1.3\%} on CoreMark-Pro and \emph{3.4\%} on BEEBS.
The highest overhead is 4.9\% from CoreMark-Pro's loops benchmark and
24.8\% from BEEBS's {\tt bubblesort} benchmark.
The {\tt bubblesort} benchmark exhibits high overhead because it
spends most of its execution in a small loop with frequent stores;
to promote these stores, {\System} adds instructions to the loop that
compute the target address.
Another BEEBS program with high overhead is {\tt levenshtein}.
The reason is that one of its functions has a variable-length array on the stack and
that function is called in a loop; {\System} promotes
the stack allocation to the heap with {\tt malloc()} and {\tt free()}.
Without this promotion, {\System} incurs 2.2\% overhead on {\tt levenshtein}.
Nearly all (8 of 9) of the CoreMark-Pro benchmarks slow down by less than 2\%,
and 5 programs have less than 1\% overhead.
For BEEBS, 24 of the 29 programs slow down by less than 5\%; 16 programs
have overhead less than 1\%.
Tables~\ref{table:coremarkpro_perf} and~\ref{table:beebs_perf} also
show that the primary source of the overhead is typically {\storetechnique},
though for some programs e.g., {\tt core} and {\tt sglib-rbtree}, the shadow stack
induces more overhead due to extensive function calls.
CFI overhead is usually negligible because our
benchmarks seldom use indirect function calls.


\paragraph{{\SystemInvert} Performance}
{\SystemInvert} greatly decreases the overhead because it only
needs to convert the single privileged store instruction in the prologue
of a function to a unprivileged one and leave all other stores unchanged.
It incurs only \emph{0.3\%} geomean overhead on CoreMark-Pro.
Seven of the 9 programs show overhead less than 0.5\%.
For BEEBS, the geometric mean overhead is 1.9\%.
When excluding the special case of {\tt levenshtein}, the average overhead
is 1.3\%.
Twenty of the 29 programs slow down by less than 1\%.
Only three programs, {\tt sglib-rbtree}, {\tt stb\_perlin}, and
{\tt trio-sscanf}, again, except {\tt levenshtein}, slow down by
over 5\%, and all of them have very frequent function calls.

\subsection{Code Size Overhead}
\label{section:results:mem}
%
%

%
%
\begin{table}[ptb]
\centering
{\sffamily
\footnotesize{
\resizebox{\columnwidth}{!}{%
\begin{tabular}{@{}lrrrrrrr@{}}
\toprule
  & {\bf Baseline} & {\bf SS} & {\bf {\StoreAbbreviation}} & {\bf CFI} & {\bf Silhou-} & {\bf Invert} & {\bf {\SSFI}} \\
  & {\bf (bytes)} & {\bf ($\times$)} & {\bf ($\times$)} & {\bf ($\times$)} & {\bf ette ($\times$)} & {\bf ($\times$)} & {\bf ($\times$)}\\
\midrule
  {\bf Min} & 51,516 & 1.005 & 1.028 & 1.002 & 1.036 & 1.008 & 1.071 \\
  {\bf Max} & 99,156 & 1.017 & 1.111 & 1.094 & 1.193 & 1.113 & 1.315 \\
  {\bf Geomean} & --- & 1.008 & 1.068 & 1.012 & 1.089 & 1.022 & 1.172 \\
\bottomrule
\end{tabular}
}}}
\caption{Code Size Overhead on CoreMark-Pro}
\label{table:coremarkpro_codesize}
\end{table}

Small code size is critical for embedded systems with limited memory.
We therefore measured the code size overhead incurred by {\System} by measuring
the code size of the CoreMark-Pro and BEEBS benchmarks.
Due to space limitations, we only show the highest, lowest, and average
code size increases in Tables~\ref{table:coremarkpro_codesize}
and~\ref{table:beebs_codesize}.
In summary, {\System} incurs a geometric mean of 8.9\% and 2.3\%
code size overhead on CoreMark-Pro and BEEBS, respectively.

For {\System}, most of the code size overhead comes from {\storetechnique}.
As Section~\ref{section:impl:store} explains,
{\System} transforms some regular store instructions
into a sequence of multiple instructions.  Floating-point stores and
stores that write multiple registers to contiguous memory locations
bloat the code size most.  In BEEBS, {\tt picojpeg}
incurs the highest code size overhead because an unrolled loop
contains many such store instructions, and the function
that contains the loop is inlined multiple times.
For {\SystemInvert}, because it leaves nearly all stores unchanged,
its code size overhead is only 2.2\% on CoreMark-Pro
and 0.5\% on BEEBS.

\begin{table}[ptb]
\centering
{\sffamily
\footnotesize{
\resizebox{\columnwidth}{!}{%
\begin{tabular}{@{}lrrrrrrr@{}}
\toprule
  & {\bf Baseline} & {\bf SS} & {\bf {\StoreAbbreviation}} & {\bf CFI} & {\bf Silhou-} & {\bf Invert} & {\bf {\SSFI}} \\
  & {\bf (bytes)} & {\bf ($\times$)} & {\bf ($\times$)} & {\bf ($\times$)} & {\bf ette ($\times$)} & {\bf ($\times$)} & {\bf ($\times$)}\\
\midrule
  {\bf Min} & 30,144 & 1.003 & 1.005 & 1.000 & 1.009 & 1.000 & 1.009 \\
  {\bf Max} & 46,108 & 1.006 & 1.061 & 1.013 & 1.068 & 1.019 & 1.201 \\
  {\bf Geomean} & --- & 1.004 & 1.018 & 1.001 & 1.023 & 1.005 & 1.044 \\
\bottomrule
\end{tabular}
}}}
\caption{Code Size Overhead on BEEBS}
\label{table:beebs_codesize}
\end{table}

\subsection{{\StoreTechnique} vs. SFI}
\label{section:results:sfi}

An alternative to using {\storetechnique} to protect the shadow stack is
to use Software Fault Isolation (SFI)~\cite{SFI:SOSP93}.  To compare
the performance and code size overhead of {\storetechnique} against SFI,
we built a system that provides the same protections as {\System} but
that uses SFI in place of {\storetechnique}.  We dub this system
\emph{{\System}-SFI} (\emph{{\SSFI}}).
Our SFI pass instruments all store instructions
within a program other than those introduced by the shadow stack
pass and those in the HAL. Specifically, our SSFI prototype adds the
same {\tt BIC}~\cite{ARMv7M} (bitmasking) instructions as what {\System}
does for Store-Exclusives (discussed in Section~\ref{section:impl:store})
before each store to restrict them from writing to the shadow stack.


{\SSFI} incurs much higher performance and code size overhead compared
to {\System}. On CoreMark-Pro, {\SSFI} incurs
a geometric mean of 2.2\% performance overhead, nearly doubling {\System}'s
average overhead of 1.3\%; on BEEBS, {\SSFI} slows down programs by 10.2\%,
three times of {\System}'s 3.4\%.
Only on one program, the loops benchmark in CoreMark-Pro,
{\SSFI} performs better than {\System}.
For code size, {\SSFI} incurs an average of
17.2\% overhead on CoreMark-Pro and 4.4\% on BEEBS; the highest overhead
is 31.5\% and 20.1\%, respectively, while on {\System} it is 19.3\% and 6.8\%.
The specific implementation of SFI will vary on different devices due to
different address space mappings, so it is possible to get
different overhead on different boards for the same program.
In contrast, {\System}'s performance overhead on the same program
should be more predictable across different boards because the instructions
added and replaced by {\System} remain the same.

\subsection{Comparison with RECFISH and $\mu$RAI}
\label{section:results:compare}

RECFISH~\cite{RECFISH:ECRTS19} and {\uRAI}~\cite{uRAI:NDSS20} are both recently
published defenses that provide security guarantees similar to {\System} but
via significantly different techniques.  Like
{\System}, they provide return address integrity coupled with
coarse-grained CFI protections for ARM embedded
architectures.  As each
defense has distinct strengths and weaknesses, the choice of
defense depends on the specific application to be protected.
To compare {\System} with RECFISH and {\uRAI} more directly and fairly,
we also evaluated {\System} with BEEBS and the
original CoreMark benchmark using only SRAM and present their
performance numbers.

RECFISH~\cite{RECFISH:ECRTS19}, which is designed for real-time systems,
runs code in unprivileged mode and uses supervisor calls to privileged code
to update the shadow stack.  Due to
frequent context switching between privilege levels, RECFISH incurs
higher overhead than {\System} or {\uRAI}. For the 24 BEEBS benchmarks that
RECFISH and
{\System} have in common,%
\footnote{We obtained RECFISH's detailed performance data
on BEEBS via direct correspondence with the RECFISH authors.}
RECFISH incurs a geometric mean of 21\% performance
overhead, and approximately 30\% on CoreMark whereas {\System} incurs
just 3.6\% and 6.7\%, respectively.
Unlike the other two defenses, RECFISH patches binaries;
no application source code or changes to the compiler are needed.


{\uRAI}~\cite{uRAI:NDSS20} protects return addresses, in part, by encoding them
into a single reserved register and guaranteeing this register is never
corrupted. This approach is more complicated but requires no protected
shadow stack.  Consequently, {\uRAI} is very efficient
for most function calls, incurring three to five cycles for each call-return.
However, there are cases,
such as calling a function from an uninstrumented library, when {\uRAI} needs
to switch hardware privilege levels to save/load the reserved register
to/from a safe region, which is expensive.


The {\uRAI} paper~\cite{uRAI:NDSS20} reports an average of 0.1\% performance
overhead on CoreMark
and five IoT applications. The {\uRAI} authors observed that one IoT program,
{\tt FatFs\_RAM}, saw a 8.5\% speedup because their transformation triggered
the compiler to do a special optimization that was not performed on the
baseline code.  When accounting for this optimization, {\uRAI} incurred an
overhead of 6.9\% on {\tt FatFs\_RAM} and 2.6\% on average for all benchmarks.
We measured the performance of CoreMark
using {\System}; the result is 6.7\% overhead compared to {\uRAI}'s
reported 8.1\%~\cite{uRAI:NDSS20}.

Finally, we observe that {\System}'s {\storetechnique} is a general technique
for intra-address space isolation. Thus, {\System} can be extended to protect
other security-critical data in memory, which Section~\ref{section:extend}
discusses.  In contrast, {\uRAI} only
protects a small amount of data by storing it within a reserved
register; its approach cannot be as easily extended to protect
arbitrary amounts of data.  {\uRAI}
does rely on SFI-based instrumentation in exception handlers for memory
isolation, but our results in Section~\ref{section:results:sfi} show that
{\storetechnique} is more efficient than SFI and could therefore be
used to replace SFI in {\uRAI}.

\section{Extensibility}
\label{section:extend}

Although {\System} focuses on providing control-flow and return address
integrity for bare-metal applications, it can also be extended to other use
cases.  For example, with minimal modification, {\System} can be used to
protect other security-critical data in memory, such as CPI's sensitive pointer
store~\cite{CPI:OSDI14} or the kernel data structures within an embedded OS
like Amazon FreeRTOS~\cite{FreeRTOS:Amazon}. 
  
With moderate modification, {\System} can also emulate the behavior of running
application code in unprivileged mode on an embedded OS.   First, the kernel of
the embedded OS would need to configure the MPU to disable unprivileged write
access to all kernel data.  Second, the embedded OS kernel's scheduler
would need to disable unprivileged write access to memory of background
applications.  
Third, in addition to {\storetechnique}, {\System} would need to
transform loads in the application code into unprivileged loads in order to
protect the confidentiality of OS kernel data structures.  It would
also need to ensure that the embedded OS kernel code contains no CFI
labels used by user-space applications.  Fourth, the
privileged code scanner must be adjusted to forbid all privileged instructions
(as opposed to only those that can be used to bypass {\System}'s protections)
in application code, forbid direct function calls to internal functions of the
kernel, and allow privileged instructions in the embedded OS kernel. Fifth,
since the stack pointer of background applications needs to be spilled to
memory during context switch, the embedded OS kernel must protect the stack
pointer of applications from corruption in order to enforce {\System}'s
security guarantee of return address integrity. One simple solution would be
storing application stack pointers to a kernel data structure not writable by
application code.  Finally, system calls require no changes. 
In {\ISA}~\cite{ARMv7M}, application code calls a system call using the
{\tt SVC} instruction, which generates a supervisor call exception.
A pointer to the exception handler table
(which stores the address of exception handler functions)
is stored in a privileged register within the {\tt System} region;
{\System} can protect both the {\tt System} region and the exception
handler table to ensure that the {\tt SVC} instruction always
transfers control to a valid system call entry point.
Also, regardless of current privilege mode, exception handlers in
{\ISA}, including the supervisor call handler, will execute in
privileged mode and switch the stack pointer to use the
kernel stack~\cite{ARMv7M}.
Therefore, system calls require no change for {\System} to work as intended.
%

\section{Related Work}
\label{section:related}

%
%
%

\paragraph{Control-Flow Hijacking Defenses for Embedded Systems}
Besides RECFISH~\cite{RECFISH:ECRTS19} and {\uRAI}~\cite{uRAI:NDSS20},
which Section~\ref{section:results:compare} discusses,
there are several other control-flow hijacking defenses for embedded devices.
CFI CaRE~\cite{CFICaRE:RAID17} uses supervisor calls and TrustZone-M
technology, available on the ARMv8-M~\cite{ARMv8M} architecture
but not on ARMv7-M, to provide coarse-grained CFI and a protected shadow stack.
%
CFI CaRE's performance overhead on CoreMark is 513\%.
SCFP~\cite{SCFP:EuroOakland18} provides fine-grained CFI by extending
the RISC-V architecture. Unlike {\System}, SCFP is a pure CFI defense and
does not provide a shadow stack. Therefore, it cannot mitigate attacks such
as control-flow bending~\cite{CFBending:USS15} while {\System} can, as
Section~\ref{section:security:attack_surface} shows.


\paragraph{Use of Unprivileged Loads/Stores}
Others~\cite{uXOM:USS19,ILDI:DAC17} have explored the use of
ARM's unprivileged loads and stores to provide security guarantees; however,
these works differ from \System's {\storetechnique} in both implementation and
application.
{\uXOM}~\cite{uXOM:USS19} transforms regular load instructions to unprivileged
ones to implement execute-only memory on embedded systems.
Aside from differences in the provided security guarantees---i.e.,
execute-only memory versus control-flow and return address integrity---these
systems differ in how they handle dangerous instructions that could be
manipulated to bypass protections. In particular, {\uXOM} inserts
verification routines before unconverted load/store instructions to ensure
that they will not access security-critical memory regions
while {\System} leverages CFI and other forward branch protections
to prevent unexpected instructions from being executed.
ILDI~\cite{ILDI:DAC17} combines unprivileged loads and stores on the ARMv8-A
architecture along with the PAN state and {\tt hyp mode} to isolate data within
the Linux kernel---the latter two features are not available on the {\ISA}
systems targeted by \System.


\paragraph{Intra-Address Space Isolation}
{\System} protects the shadow stack by leveraging {\storetechnique}.
Previous work has explored other methods of intra-address space isolation
which could be used to protect the shadow stack.  Our evaluation in
Section~\ref{section:results:sfi} compares {\System} to
Software Fault Isolation (SFI)~\cite{SFI:SOSP93}, so we focus on other
approaches here.
%
%



ARM Mbed $\mu$Visor~\cite{ARMuVisor}, MINION~\cite{MINION:NDSS18}, and
ACES~\cite{ACES:USS18} enforce memory compartmentalization on embedded
systems using the MPU.  They all dynamically reconfigure the MPU at
runtime but target different scenarios; Mbed $\mu$Visor and MINION
isolate processes from each other at context switches, and ACES dissects
a bare-metal application at function boundaries for intra-application
isolation.  As discussed previously, isolation that requires protection
domain switching is poorly-suited to security instrumentation that requires
frequent crossing of the isolation boundaries---such as {\System}'s shadow stack
accesses.

ARMlock~\cite{ARMlock:CCS14} uses ARM domains to place pages into different
protection domains; a privileged register controls access to pages
belonging to different domains.  ARM domains are only available for
CPUs with MMUs~\cite{ARMv7AR,ARMv7M} and therefore cannot be
used in {\ISA} systems.  Additionally, access to ARM domains can only
be modified in privileged mode; software running in user-space must
context switch to privileged mode to make changes.


\paragraph{Information Hiding}  
Given the traditionally high cost of intra-address space isolation, many
defenses hide security-critical data by placing
it at a randomly chosen  address. This class of techniques
is generally referred to as information hiding.  For example,
EPOXY~\cite{EPOXY:Oakland17} includes a backward-edge control-flow hijacking
defense that draws inspiration from CPI~\cite{CPI:OSDI14}---relying on
information hiding to protect security-critical data stored in memory.
Consequently, an adversary with a write-what-where vulnerability (as assumed in
our threat model) can bypass EPOXY protections.


Fundamentally, information hiding is unlikely to be a strong defense on
embedded systems as such systems tend to use only a fraction of the  address
space (and the memory is directly mapped)  which limits the entropy attainable.
For example, our evaluation board only has 2~MB of memory for code; if each
instruction occupies two bytes, randomizing the code segment provides at most
20 bits of entropy.  In contrast, {\System}'s defenses are effective even if
the adversary has full knowledge of the memory layout and contents.


\paragraph{Memory Safety}
Memory safety provides strong protection but incurs high overhead.  Solutions
using shadow
memory~\cite{ASan:ATC12,BBC:USS09,BBAC:ISSREW12,PAMD:ISMM17,WIT:Oakland08}
may consume too much memory for embedded systems.  Other
solutions~\cite{JonesKelly:AADEBUG97,CCured:POPL02,%
CRED:NDSS04,SAFECode:ICSE06,SAFECode:PLDI06} incur too much performance
overhead.
nesCheck~\cite{nesCheck:AsiaCCS17} is a memory safety compiler for
TinyOS~\cite{TinyOS:ASPLOS00} applications which induces 6.3\% performance
overhead on average.  However, nesCheck cannot support binary code libraries
as it adds additional arguments to functions.
Furthermore, nesCheck's performance relies heavily on static analysis.
We believe that, due to their simplicity, the benchmarks used in the
nesCheck evaluation are more amenable to static analysis than applications
for slightly more powerful embedded systems (such as ours).
In contrast, {\System}'s performance does not depend on static analysis's
precision.

\section{Conclusions and Future Work}
\label{section:conclude}


In conclusion, we presented {\System}: a software control-flow hijacking
defense that guarantees the integrity of return addresses for embedded systems.
To minimize overhead, we proposed {\SystemInvert}, a system which provides
the same protections as {\System} with significantly lower overhead at
the cost of a minor hardware change.
We implemented our prototypes for an {\ISA} development board.
Our evaluation shows that {\System} incurs low performance overhead:
a geometric mean of 1.3\% and 3.4\% on two benchmark suites, and
{\SystemInvert} reduces the overhead to 0.3\% and 1.9\%.
We are in the process of opening source the {\System} compiler
and related development tools. They should be available at
\url{https://github.com/URSec/Silhouette}.

We see two primary directions for future work.
First, we can optimize {\System}'s performance.
For example, Section~\ref{section:security:ss_integrity} shows that
{\System} ensures that the stack pointer stays within the stack region.
Consequently, store instructions using the {\tt sp} register and an immediate to
compute target addresses are unexploitable;
{\System} could elide {\storetechnique} on such stores.
Second, we can use {\System} to protect other memory
structures, such as the safe region used in CPI~\cite{CPI:OSDI14} and
the process state saved on interrupts and
context switches (like previous work~\cite{KCoFI:Oakland14} does).

\section*{Acknowledgements}
\label{section:ack}

The authors thank the anonymous reviewers for their insightful
comments and Trent Jaeger, our shepherd, for helping us improve our paper.
This work was funded by NSF awards CNS-1618213 and CNS-1629770 and
ONR Award N00014-17-1-2996.

{\small
\bibliographystyle{plain}
\bibliography{silhouette}
}

\appendix
\section{Design to Support {\tt setjmp}/{\tt longjmp}}
\label{section:setjmp}

Calls to {\tt setjmp} and {\tt longjmp} can undermine {\System}'s
return addresses integrity guarantees because {\tt longjmp} uses a
return address from its {\tt jmp\_buf}
argument which could be located in corruptible global,
heap, or stack memory.  Applications might also misuse {\tt setjmp} and
{\tt longjmp}, such as calling {\tt longjmp} after the function that
called {\tt setjmp} with the corresponding {\tt jmp\_buf} returns, leading to
undefined behaviors exploitable by attackers.
{\System} modifies the implementation of {\tt setjmp} and {\tt
longjmp} to support them while maintaining its return address
integrity guarantees.

\begin{figure}[b]
  \centering
  \includegraphics[scale=0.40]{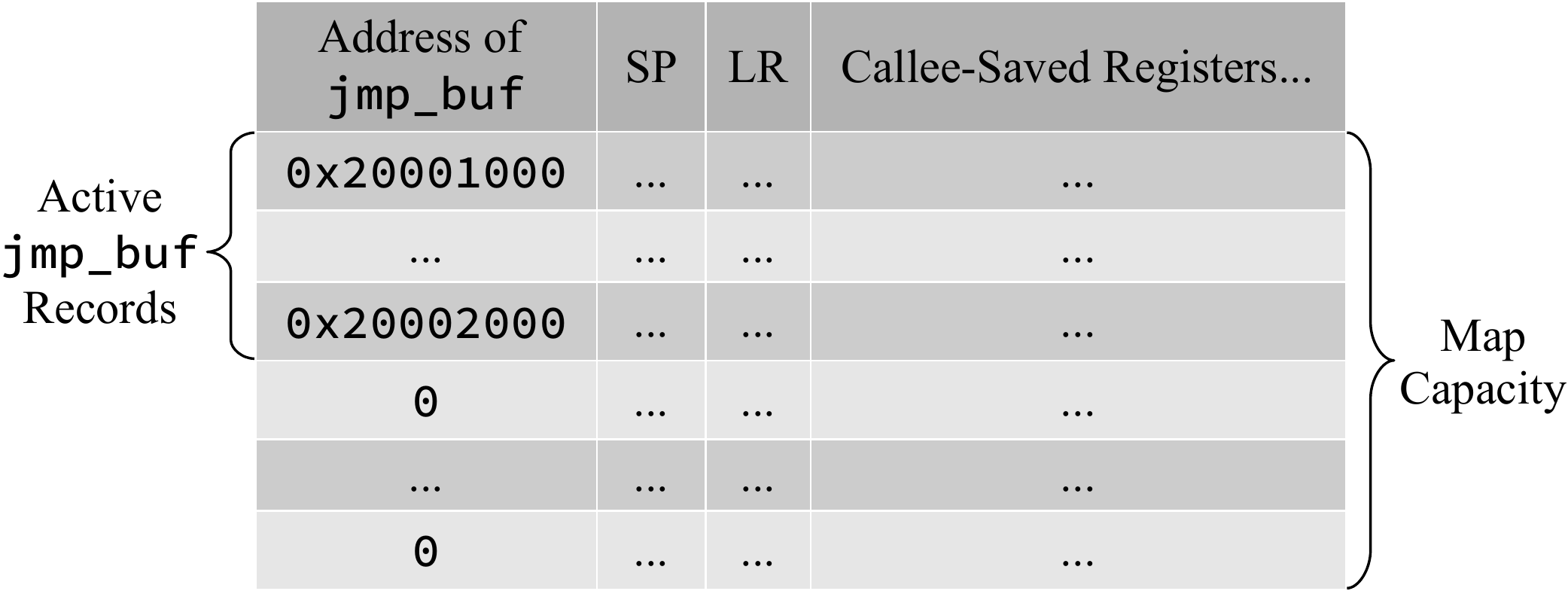}
  \caption{Format of {\tt jmp\_buf} Records}
  \label{fig:jmpbuf-format}
\end{figure}

\begin{algorithm}[ptb]
  \SetKwData{Buf}{buf}
  \SetKwData{Map}{map}
  \SetKwData{SP}{sp}
  \SetKwData{LR}{lr}
  \SetKwData{EntryI}{e}
  \SetKwData{Size}{size}
  \SetKwData{Capacity}{capacity}
  \SetKwFunction{Error}{Error}
  \SetKw{In}{in}
  \SetProgSty{}

  \KwIn{A {\tt jmp\_buf} \Buf}

  \ForEach{entry \EntryI \In \Map}{
    \If{\EntryI.\Buf $==$ \&\Buf}{
      \EntryI.\{\SP, \LR, \dots\} $\gets$ \{\SP, \LR, \dots\}\;
      \KwRet{0}\;
    }
  }
  \eIf{\Map.\Size $<$ \Map.\Capacity}{
    Insert a new entry \{\&\Buf, \SP, \LR, \dots\} into \Map\;
    \Map.\Size $\gets$ \Map.\Size $+1$\;
    \KwRet{0}\;
  }{
    \Error{``Map reached its capacity''}\;
  }

  \caption{{\System} {\tt setjmp}}
  \label{alg:setjmp}
\end{algorithm}

\begin{algorithm}[ptb]
  \SetKwData{Buf}{buf}
  \SetKwData{Val}{val}
  \SetKwData{Map}{map}
  \SetKwData{SP}{sp}
  \SetKwData{LR}{lr}
  \SetKwData{Entry}{buf\_entry}
  \SetKwData{EntryI}{e}
  \SetKwData{Size}{size}
  \SetKwFunction{Error}{Error}
  \SetKw{Null}{null}
  \SetKw{In}{in}
  \SetKw{Break}{break}
  \SetProgSty{}

  \KwIn{A {\tt jmp\_buf} \Buf}
  \KwIn{An integer \Val}

  \Entry $\gets$ \Null\;
  \ForEach{entry \EntryI \In \Map}{
    \If{\EntryI.\Buf $==$ \&\Buf}{
      \Entry $\gets$ \EntryI\;
      \Break\;
    }
  }
  \If{\Entry $==$ \Null}{
    \Error{``Invalid jmp\_buf''}\;
  }
  \ForEach{entry \EntryI \In \Map}{
    \If{\EntryI.\SP $<$ \Entry.\SP}{
      Invalidate \EntryI\;
      \Map.\Size $\gets$ \Map.\Size $-1$\;
    }
  }
  \{\SP, \LR, \dots\} $\gets$ \Entry.\{\SP, \LR, \dots\}\;
  \eIf{\Val $==$ 0}{
    \Return{1}\;
  }{
    \Return{\Val}\;
  }

  \caption{{\System} {\tt longjmp}}
  \label{alg:longjmp}
\end{algorithm}

Specifically, {\System} reserves part of the protected shadow stack region
to store a map of active {\tt jmp\_buf} records in use by
the program.  Figure~\ref{fig:jmpbuf-format} shows the format of a map
entry; the address of a {\tt jmp\_buf} passed to
{\tt setjmp}/{\tt longjmp} serves as a key, and all callee-saved
registers plus {\tt sp} and {\tt lr} are values.
Algorithms~\ref{alg:setjmp} and \ref{alg:longjmp} depict the design of
our custom {\tt setjmp} and {\tt longjmp}, respectively.
When the application calls {\tt setjmp}, instead of saving the
execution context to the application-specified {\tt jmp\_buf},
{\System}'s {\tt setjmp} saves it to the map by inserting a new
entry or overriding an existing entry, based on the address of
{\tt jmp\_buf}.  If we are inserting a new entry and the number of
active {\tt jmp\_buf} records reaches the map's capacity,
{\System}'s {\tt setjmp} reports an error and aborts the program;
this is not a practical problem as we expect the program to have
only a few {\tt jmp\_buf}s.  We can also provide an option for the
application developer to specify a desired size of the map.
Our {\storetechnique} pass will recognize this safe
version of {\tt setjmp} and generate regular stores (instead of
unprivileged stores) for it to access the map.  Saving the execution
context in the protected region ensures the integrity of saved stack
pointer values and return addresses.

{\System}'s {\tt longjmp} checks if the
address of the supplied {\tt jmp\_buf} matches an entry in the map.
If no matched entry is found, either
the supplied {\tt jmp\_buf} is invalid or
the supplied {\tt jmp\_buf} has expired due to function
returns or a call to {\tt longjmp} on an outer-defined
{\tt jmp\_buf} (both explained below).
In both cases, execution is aborted.
If a matched entry is found, {\System}'s {\tt longjmp}
first invalidates all
entries in the map that have a smaller {\tt sp} value than that of
the matched entry; these {\tt jmp\_buf}s
become expired when the control flow is unwound to an outer call
site of {\tt setjmp}.  The execution context stored in the matched entry
is then recovered.

The remaining case is that, when a function that calls
{\tt setjmp} returns, the {\tt jmp\_buf}s used in the function and in
its callees become obsolete.  {\System} handles this case by inserting
code in the epilogue of such functions to invalidate all the map entries
whose {\tt sp} value is smaller than or equal to the current {\tt sp}
value.  This ensures that future calls to {\tt longjmp} do not use
obsolete {\tt sp} and {\tt lr} values.

\end{document}